\documentclass[aps,twocolumn,prd,superscriptaddress,floatfix,nofootinbib,longbibliography]{revtex4-2}
\usepackage{graphicx,longtable,mathrsfs,color,array}
\usepackage[dvipsnames]{xcolor} 
\usepackage{adjustbox}
\usepackage{tikz}
\usepackage{amssymb,amsmath,mathtools,mathrsfs,slashed,bm} 
\usepackage{braket}
\usepackage{epsfig,subfigure,placeins,float} 
\usepackage{booktabs,longtable,multirow} 
\usepackage{dsfont}
\usepackage{ulem}
\usepackage{color}
\usepackage[qm]{qcircuit}
\usepackage{placeins}
\usepackage{comment}
\usepackage{siunitx}
\usepackage{todonotes}
\usepackage[pdftex,
            pdftitle={Barren-plateau free variational quantum simulation of Z2 LGTs},
            pdfauthor={Fariha Azad, Matteo Inajetovic, Stefan Kühn, Anna Pappa}
            bookmarks,
            colorlinks,
            menucolor=black,
            plainpages=false,
            pdfpagelabels,
            hypertexnames=false]{hyperref}

\begin{document}
\title{Barren-plateau free variational quantum simulation of \texorpdfstring{$\mathbb{Z}_2$}{Z2} lattice gauge theories}

\author{Fariha Azad}
\affiliation{Electrical Engineering and Computer Science Department, Technische Universität Berlin, 10587 Berlin, Germany}
\author{Matteo Inajetovic}
\affiliation{Electrical Engineering and Computer Science Department, Technische Universität Berlin, 10587 Berlin, Germany}
\author{Stefan Kühn}
\affiliation{Deutsches Elektronen-Synchrotron DESY, Platanenallee 6, 15738 Zeuthen, Germany}
\author{Anna Pappa}
\affiliation{Electrical Engineering and Computer Science Department, Technische Universität Berlin, 10587 Berlin, Germany}

\date{\today}

\begin{abstract}
    In this work, we implement a variational quantum eigensolver (VQE) suitable for investigating ground states and static string breaking in a $\mathbb{Z}_2$ lattice gauge theory (LGT).
    We consider a two-leg ladder lattice with Kogut-Susskind staggered fermions and verify the results of the VQE simulations using tensor network methods. 
    We find that for varying Hamiltonian parameter regimes and in the presence of external charges, the VQE is able to arrive at the gauge-invariant ground state without explicitly enforcing gauge invariance through penalty terms. Additionally, experiments showing string breaking are performed on IBM's quantum platform. 
    Thus, VQEs are seen to be a promising tool for $\mathbb{Z}_2$ LGTs, and 
    could serve as a stepping stone toward studies of other gauge groups.
    We find that the scaling of gradients with the number of qubits is favorable for avoiding barren plateaus.  
    Furthermore, strategies that avoid barren plateaus arise naturally as features of LGTs, such as choosing the initialization by setting the Gauss law sector and restricting the Hilbert space to the gauge-invariant subspace. 
\end{abstract}

\maketitle

\section{Introduction}

Gauge theories, i.e.\ field theories with a local symmetry, are the theoretical foundation for the description of particle physics. At the same time, analytical access in the non-perturbative regime is often not possible. One way forward is lattice gauge theory (LGT), in which space-time is discretized on a lattice~\cite{Wilson1974}. 
In this framework, matter fields are assigned to the lattice sites, while gauge fields are associated with the connecting links, all while maintaining gauge symmetry. In particular, this discretization allows for a numerical approach with sophisticated Monte Carlo methods.

While Monte Carlo methods have demonstrated excellent predictive power for static problems, they are limited by the sign problem in certain parameter regimes, leaving many relevant questions unanswered.  
This has triggered the ongoing effort to develop quantum and quantum-inspired methods for LGTs, such as tensor networks (TNs), to overcome these limitations. 
Current quantum processors are a long way from being used for realistic simulations of gauge theories.
Successful demonstrations have been carried out for low-dimensional toy models~\cite{Martinez2016,Atas2021,Atas2022,farrell2024scalable, farrell2024quantum,Klco2018,Klco2019, Kokail2018, meth2025simulating,Cochran2024,Cobos2025, ciavarella2021trailhead, ciavarella2024quantum} of confinement~\cite{Mildenberger2025,Ciavarella2025,alexandrou2025realizingstringbreakingdynamics,Crippa2024a, gonzalez2025observation}, and scattering dynamics ~\cite{Davoudi2024,Zemlevskiy2024,chai_fermionic_2025, chai_resource-efficient_2025, davoudi2025quantum, schuhmacher2025observation}.
For reviews, see, e.g.\ Refs.~\cite{Banuls2019,Banuls2020,Funcke2023a,bauer2024quantum, Halimeh2025}.

In this work, we 
{apply a variational quantum eigensolver (VQE) to} a $\mathbb{Z}_2$ LGT~\cite{Horn1979} on a two-leg ladder geometry coupled to Kogut-Susskind fermions~\cite{Kogut1975}.

{In so doing, we are able to recover ground
state properties and to explore static string breaking}.
The two-leg ladder lattice is an intermediate step between one and two dimensions. It is the simplest geometry that features magnetic plaquette interactions, and therefore all of the terms of the LGT Hamiltonian, while still being efficiently verifiable using tensor network techniques. 
The simplicity of the $\mathbb{Z}_2$ LGT is amenable to noisy intermediate-scale quantum (NISQ) devices, while still exhibiting complex phenomena, such as confinement.

Confinement is the mechanism that explains why isolated quarks or gluons are not observed in nature. 
The dynamical process of string breaking is fundamental to confining gauge theories, and is described by quantum chromodynamics. 
Using $\mathbb{Z}_2$ LGT as a toy model, we investigate static aspects of string breaking as a consequence of confinement. 
At large values of the coupling, when two charges are confined in the $\mathbb{Z}_2$ LGT, by Gauss law they are necessarily connected by an electric flux tube.
Increasing their distance leads to a linearly rising potential. At large distances, it becomes energetically favorable to break the flux string, and here this is done by forming an additional pair of charges. In addition, at intermediate values of the coupling, a roughening transition can occur, which is a phase transition where the confining flux tube becomes rough, i.e., its transverse fluctuations become significant. This transition has recently been observed in TN simulations~\cite{marcantonio_roughening_2025}.

{Here we study this problem using VQEs.}
VQEs are hybrid quantum-classical algorithms for estimating expectation values 
and are widely studied in the NISQ era as they do not require error-corrected qubits. 
They are among numerous quantum approaches to study LGTs in the NISQ era \cite{irmejs2023quantum, nagano2023quench, cobos2024noise, balaji2026perturbation}. 
However, they have no provable guarantees, so the question remains: are there problems that variational quantum algorithms are better suited to solve compared to classical methods? 

While we do not yet have the quantum hardware to answer this question, recent investigations of variational quantum algorithms have unearthed many issues. 
In particular, there seems to be a connection between the trainability of a VQE (the absence of \textit{barren plateaus}) and efficient classical simulation~\cite{cerezo2024doesprovableabsencebarren}. 
Thus, it is important that proposals involving barren-plateau free VQEs consider the scaling with system size and the effects of noise.

We observe features within the framework of $\mathbb{Z}_2$ LGT, and LGTs in general, that are compatible with common strategies for avoiding barren plateaus. 
Warm starts are one such strategy, where one 
{initializes} close to a solution, where there may be narrow gorges~\cite{Puig2025} in the search landscape. 
Analogously, for investigations of LGT, it is necessary to initialize to some Gauss law sector, 
{which thereby restricts Hilbert space into a direct sum structure of superselection sectors \cite{panizza2022entanglement}.}
The search process for a VQE is benefited by constraining to a subspace of Hilbert space~\cite{Nakanishi19};
by gauge symmetry for LGTs, this is the subspace of the gauge-invariant, physical, states.

We draw upon these features using two types of ansatz circuits.
One is restricted entirely to the symmetric subspace of physical states by being composed of gauge invariant terms. 
The other has overlap with states outside of the physical subspace, but nonetheless 
{learns the correct gauge degrees of freedom. 
Thus it is not necessary to impose gauge invariance by use of penalty terms, that can generally impede training.}
For the simulations we consider scaling with system size and numbers of shots, to find good convergence of ground states and can witness static string breaking. 
Results are also obtained using IBM's quantum hardware, where we perform experiments in the region of the transition into the string-broken phase. 
These findings underscore the usefulness and suitability of VQEs in the investigation of LGTs.

\section{Background and methods}

\subsection{Lattice gauge theories}

\begin{figure}
    \centering
    \includegraphics[width=\linewidth]{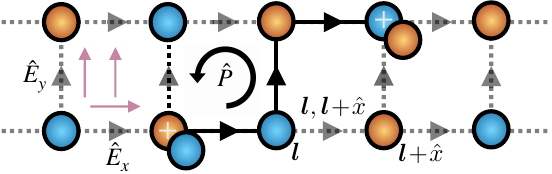}
    \caption{\textit{Features of LGTs on the two-leg ladder lattice.} The LGT specifies matter sites and links between them pertaining to the underlying gauge field.
    The arrows on the links show the orientation of the lattice.
    We have labeled $\hat E_x$ and $\hat E_y$ the orientations of the electric field and the ring labeled $\hat P$ indicates the magnetic plaquette term. 
    For the vertex $\bm l$, we have indicated the notation for the adjacent link and matter site that is one unit vector $\hat x$ across. 
    The red arrows show how we linearly arrange the qubits to perform the Jordan-Wigner transformation: we count sites from the leftmost lower corner, then traverse upwards, then rightwards along the lattice.
    The alternating colors of the matter sites illustrate the coupling to staggered fermions where the sign alternates. 
    On this diagram we have also included an example string configuration, through the solid arrow connecting two charges that have been placed. 
    The overlapping circles show the change in sign, indicating the placed charges. 
    }
    \label{fig:lgt-diagram}
\end{figure}

The defining feature of gauge theories is invariance under local transformations. 
This leads to the presence of a local conservation law or a continuity equation that determines the interplay between the matter and gauge degrees of freedom. In the Hamiltonian formulation using temporal gauge, this results in an additional constraint known as Gauss law~\cite{Kogut1975}.

Lattice gauge theories (LGTs) provide a means to probe gauge theories numerically. 
Numerical studies such as those using Monte Carlo techniques are limited by the sign problem in certain parameter regimes, thus a long-term goal for quantum computing is to access these intractable regions. 
One way to discretize gauge theory considers placing matter sites on the vertices and the underlying gauge group acts on the connecting links.

The pure gauge part of the theory is constructed following Ref.~\cite{Horn1979} and the lattice notation for the two-dimensional ladder geometry we consider is illustrated in Fig.~\ref{fig:lgt-diagram}.
When the underlying gauge group is $\mathbb{Z}_2$, the unitary operators acting on the links fulfill the $\mathbb{Z}_2$ algebra, so they can be re-written in terms of the Pauli matrices $X$ and $Z$.
For the ladder geometry, the resulting pure gauge part of the Hamiltonian is given by: 
\begin{equation}
    \begin{split}
        H_g = &- \mu \sideset{}{_{\bm{l},k=x,y}}{\sum} X_{\bm{l},\bm{l}+\hat{k}} \\
            &- \sideset{}{_{\bm{l}}}{\sum} Z_{\bm{l}, \bm{l}+\hat{x}}Z_{\bm{l}, \bm{l}+\hat{y}}Z_{\bm{l}+\hat{x}, \bm{l}+\hat{x}+\hat{y}}Z_{\bm{l}+\hat{y}, \bm{l}+\hat{x}+\hat{y}},
            \label{eq:H_g}
    \end{split}
\end{equation}
where the operators act on the links emanating from the matter site with coordinate $\bm l = (l_x,l_y)\in\mathbb{Z}^2$, and $\hat{x}$, $\hat{y}$ are unit vectors in $x$ and $y$-direction.
The first part of this equation is the electric energy, where $X$ takes the role of an electric field. The second term corresponds to the magnetic energy contribution and represents a product of the $Z$ operators on a plaquette of the lattice. 
The parameter $\mu>0$ tunes the relative strength between the electric and magnetic energy. 

To couple the theory to fermionic matter, we consider Kogut-Susskind staggered fermions~\cite{Kogut1975}. This formulation essentially corresponds to separating the spinor components to different lattice sites, resulting in alternating the signs of the masses and charges across the lattice as reflected in the orange and blue coloring in Fig.~\ref{fig:lgt-diagram}.

Then the terms representing the gauge-matter interaction and the fermion mass read:
\begin{equation}
    \begin{aligned}
            H_m &= J \sideset{}{_{\bm{l},\hat k}}{\sum}
            \left(\phi_{\bm{l}}^\dagger Z_{\bm{l},\bm{l}+\hat{k}}\phi_{\bm{l}+\hat{k}} + \text{h.c.}\right) \\
            &+ m \sideset{}{_{\bm{l}}}{\sum}(-1)^{l_1+l_2} \phi^\dagger_{\bm{l}}\phi_{\bm{l}},
            \label{eq:H_m}
    \end{aligned}
\end{equation}
where $J,m>0$, $\phi_{\bm l}$ is the spinless fermionic operator at site $\bm l$, and the connections $Z_{\bm{l},\bm{l}+\hat{k}}$ lift the gauge symmetry to be local for this Hamiltonian, that otherwise consists of nearest-neighbor interactions. In the limit $J\rightarrow 0$ the gauge and matter fields do not interact, hence we refer to it as the non-interacting limit. 

The physical states satisfy Gauss law, which constrains the in-going and out-going electric flux on the links from a matter site and relates the staggered charge. 
In terms of the $\mathbb{Z}_2$ algebra, $\forall \bm{l}$ and physical states $\ket{\psi}$, $G_{\bm l}\ket{\psi}=q_{\bm l}\ket{\psi}$, where $q_{\bm l}=\pm 1$ and 
\begin{equation}
    G_{\bm{l}} = (-1)^{l_1 + l_2} (-1)^{n_l} \sideset{}{_{\hat{k}}}{\prod} X_{\bm{l},\bm{l} + \hat{k}} \sideset{}{_{\hat{k}}}{\prod}  X_{\bm{l} - \hat{k}',\bm{l}}.
    \label{eq:gauss_law}
\end{equation}
The prefactor represents the different nature of the odd and even lattice sites, and we define the number operator $n_l=\phi_l ^\dagger \phi_l$. Following Ref.~\cite{alexandrou2025realizingstringbreakingdynamics}, we interpret $q_{\bm l}= 1$ as the absence of a charge, whereas a value of $q_{\bm l}=-1$ corresponds to a static charge at site $\bm l$.
Depending on the point of the lattice, the operators acting on the links are replaced by the boundary condition $X_{\bm l, \bm l+\hat{x}}\!\rightarrow1$ (see Ref.~\cite{alexandrou2025realizingstringbreakingdynamics} for details). 

Since $\forall \bm l$, $[H,G_{\bm l}] =0$, $G_{\bm l}$ is a local symmetry that decomposes the Hilbert space into separate sectors depending on the charge. 
Taking into account the staggering, we can consider the Gauss law sector of the states we perform calculations with. 

For simulations of LGTs, it is necessary for physical results to be gauge invariant, making it necessary to enforce Gauss law.
One way to enforce gauge invariance is through penalty terms that are proportional to \eqref{eq:gauss_law}. 
Gauge-invariant states subject to gauge-invariant operators remain in this physical subspace, i.e., the symmetry remains intact.
This subspace is still exponentially large, while being non-trivial to split into further gauge-invariant subspaces
~\cite{Buividovich_2009, Tagliacozzo2014}. 

Incorporating terms proportional to \eqref{eq:gauss_law} allows for enforcing Gauss law, where infinitely massive static charges can be introduced to the theory by setting the corresponding $q_{\bm l} = -1$.  
A consequence of Gauss law is that a pair of charges placed on matter sites in vicinity of each other are necessarily connected by a flux tube or \textit{string}. 
These charges experience a linearly rising potential that is proportional to their separation.

From a certain distance onward, it becomes energetically favorable to break the string, which creates an additional pair of charges. 
This is known as static string breaking, and is depicted in Fig.~\ref{fig:lgt-diagram}, where a configuration of a string is shown between two sites with charges, indicated by the opposite coloring. 
Static string breaking and confinement has also been investigated for $\mathbb{Z}_2$ LGT in the following works~\cite{Lumia2022Z2,Xu2025, marcantonio_roughening_2025, alexandrou2025realizingstringbreakingdynamics}. 

Then, the total Hamiltonian is as follows, 
\begin{equation}
	H = H_g + H_m + V \sideset{}{_{\bm{l}}}{\sum}(G_{\bm{l}} -q_{\bm{l}})^\dagger (G_{\bm{l}} -q_{\bm{l}}),
	\label{eq:H}
\end{equation}
with $V>0$. 
Placing charges amounts to adding penalty terms on the appropriate site with the opposite sign.

To compute expectation values of this Hamiltonian on a quantum computer, we perform the Jordan-Wigner transformation \cite{Hamer1997Series} on the matter Hamiltonian \eqref{eq:H_m} that couples to fermionic degrees of freedom. 
For this we must choose a linear ordering of the fermions.
The gauge part and penalty terms already consist of Pauli operators.
In Fig.~\ref{fig:lgt-diagram}, the ordering zig-zags from the bottom left qubit to the top, then running across the rungs towards the left, as indicated by the red arrows. 
Subsequently, we translate the fermions to spins using $\phi_{\bm l} = \sideset{}{_{k<l}}{\prod}(i Z_k) \sigma_l ^-$, with $\sigma^\pm = \frac{1}{2}(X \pm iY)$ and $l$ being the linear index of the ordering of the fermionic fields as described. 

\subsection{Variational quantum eigensolver}

Variational quantum eigensolver (VQE)~\cite{Cerezo2021a} is a well-established hybrid quantum-classical algorithm for computing expectation values of some observable. 
On a quantum processor, one can make a choice of rotation and entangling gates to form an ansatz circuit $U(\bm \theta)$ that prepares families of quantum states $\ket{\psi(\bm \theta)}=U(\bm \theta)\ket{\psi_{in}}$. 
An approximation for the ground state $\ket{\psi}$ of a given observable $H$, minimizes the cost function:
\begin{equation}
    C(\bm \theta) = \bra{\psi(\bm \theta)}\!H\!\ket{\psi(\bm \theta)}.
    \label{eq:generic_cost}
\end{equation}

For observables formed of tensor products of simple Pauli operators, the quantum processor can efficiently evaluate the expectation value.
A classical optimizer then improves on this guess of parameters $\bm \theta$, thus completing this quantum-classical loop.  

The choice of ansatz must carefully balance incorporating physical information of the problem, i.e., being \textit{expressible}, with being \textit{trainable}. 
{Expressibility} is the extent to which the ensemble of unitaries generated by the ansatz contains the unitary that is `close' to that which minimizes the cost.
Then for classical gradient-based optimizers, {trainability} indicates that the cost landscape manifests large enough gradients to optimize the parameters and minimize the cost function. 
Barren plateaus occur when the gradients of the cost function (\ref{eq:generic_cost}) become exponentially suppressed with the system size. 
Thus the features we require of the VQE seem to be at odds -- deeper circuits are in general more expressible, but by construction are susceptible to barren plateaus.

Circumstantially, there seems to be the additional issue that the trainability of a VQE implies that it is classically simulable.
We can avoid barren plateaus by using ansatz circuits that explore smaller regions of Hilbert space related to the underlying problem structure. 
One way to reduce the dimensionality of the parameter space is to apply the symmetries of the system. 
Herein lies what enables efficient classical simulation -- this reduction tends to identify some polynomially large subspace~\cite{cerezo2024doesprovableabsencebarren}. 

As such, it is possible to diagnose barren plateaus from the size of the dynamical Lie algebra (DLA) 
spanned by the gate generators. 
In~\cite{Larocca2022diagnosingbarren} it is conjectured that the variance of gradients of the cost function scale inverse-polynomially with the dimension of the DLA. 
However, while a small DLA may avoid barren plateaus, a polynomial-sized DLA implies the problem is classically simulable~\cite{goh2025liealgebraicclassicalsimulationsquantum}. 

These are important considerations for ansatz design. 
A choice of ansatz that considers the available native gates, is known as a \textit{hardware efficient ansatz} (HEA). 
Whereas an ansatz based on unitaries formed of the problem Hamiltonian is known as the \textit{Hamiltonian variational ansatz} (HVA). 
For the Hamiltonian consisting of $k$-local terms, $H= H_K$, where each $H_K$ consists of commuting Pauli operators, the HVA can be written~\cite{Wecker2015, Wiersema2020} as: 
\begin{equation}
    \ket{\psi(\theta)} = {\sideset{}{_{K}}{\prod}\exp(-i \theta_K H_K)} \ket{\psi_0}.
\end{equation}

Initialization strategies to avoid barren plateaus have been proposed for both the HEA~\cite{Park:2024rim} and HVA~\cite{Park2024hamiltonian}. 
However, while warm starts and related initialization strategies show promise for VQEs, they are not guaranteed to find the global minimum.
Nor does there exist a general approach in subspaces containing the solution.
This occurs for only a vanishing fraction of problem instances~\cite{Puig2025}. 

In \cite{Zhang2024Absence} it is proven that local Hamiltonians training finite or logarithmic local-depth circuits (FLDC) will not encounter barren plateaus. 
Here the two-dimensional Toric code is considered, which is a variant of the pure gauge part of the $\mathbb{Z}_2$ LGT.
In general FLDC cannot be classically simulated for two dimensions and higher.

VQEs are considered promising NISQ era algorithms since they do not require error correction, however they have no provable guarantees. 
To ask what problems can test VQEs at scale, invokes questions of trainability, so we know more and more the limitations of VQEs. 
It is then crucial to identify problems that are average-case hard to solve classically while being barren-plateau free. 

\begin{figure}
    \centering
    \includegraphics[width=\linewidth]{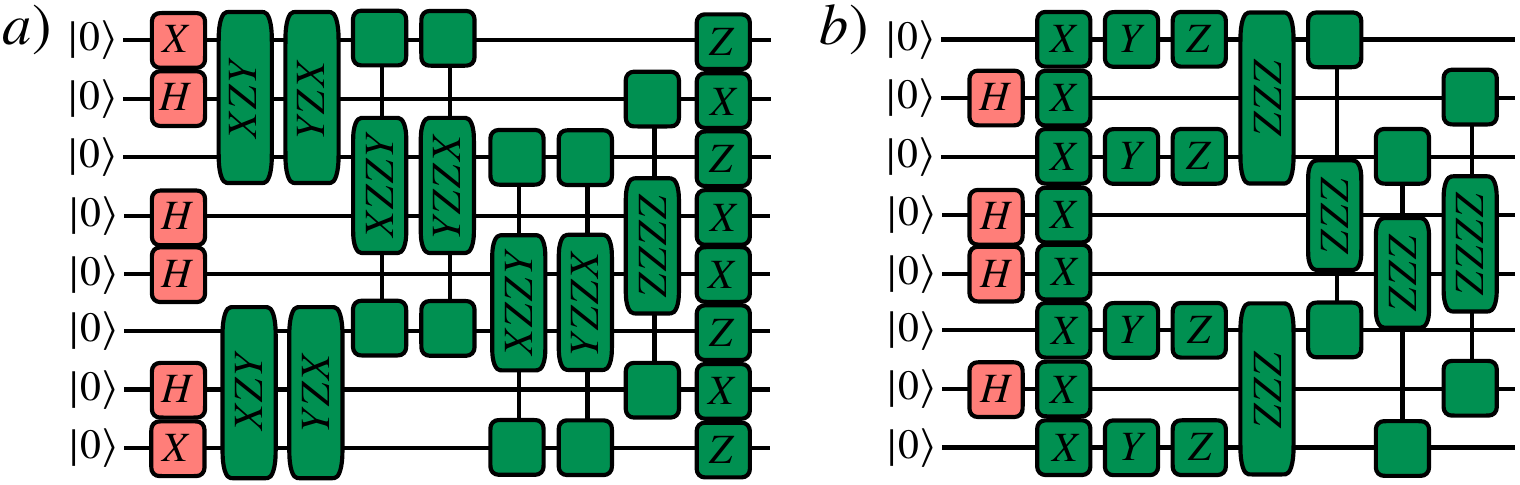}
    \caption{\textit{The ansatz circuits considered in this work}. 
    These are circuits for the simplest example of one plaquette. 
    The red gates set the initial state, while the green gates are parametrized, and form the repeating layers. Here we use the shorthand $X\rightarrow R_X(\theta)$, so each green gate has associated one variable angle. 
    a) shows the gauge-invariant (GI) ansatz  and b) the hardware-efficient, MBQC-inspired (ZZ) ansatz. }
    \label{fig:ansatz}
\end{figure}

\subsection{Tensor network states}

In order to be able to access larger system sizes beyond what can be straightforwardly done with exact diagonalization, we use numerical methods based on Matrix Product States (MPS), a particular kind of one-dimensional tensor network. Given a lattice system with local, $d$-dimensional bases, $\{\ket{i_k}\}_{i_k=1}^d$, the MPS ansatz for the wave function $\ket{\psi}$ of a system with open boundaries is given by~\cite{Orus2014a,Bridgeman2017}:
\begin{align}
    \ket{\psi} = \sum_{i_1,\dots,i_N=1}^d M_1^{i_1} M_2^{i_2} \dots M_N^{i_N}\ket{i_1}\otimes\dots \otimes\ket{i_N}.
    \label{eq:MPS_decomposition}
\end{align}

In the expression above, the $M_k^{i_k}$ are complex $\chi \times \chi$ matrices for $1<k<N$ and $M_1^{i_1}$ ($M_N^{i_N}$) is a $\chi$-dimensional row (column) vector. Thus, for a given set of indices $i_k$, the coefficients of the wave function are parametrized by a product of matrices, hence the name of the ansatz. The matrix size $\chi$, commonly referred to as the bond dimension of the MPS, determines the number of parameters in the ansatz and limits the maximum amount of entanglement that can be present in the state. 

MPS can be used as a family of variational ansätze to compute the ground state of a (local) Hamiltonian. To this end, the energy expectation value is variationally minimized by successively updating a tensor while keeping the others fixed. At each step, the optimal tensor is found by determining the ground state of an effective Hamiltonian describing the interaction of the site with its environment. Iterating this procedure until convergence, the final MPS provides an approximation for the ground state of the Hamiltonian. 
A detailed review on MPS methods can be found in Ref.~\cite{Schollwoeck2011}.

Note that in order to apply the MPS approach to our ladder geometry, we have to map the local degrees of freedom sitting on the vertices and links to a linear chain. This does not create interactions whose range scales with the system size, so the resulting one-dimensional mapping can generally be addressed efficiently with MPS.

\begin{figure}
    \centering
    \includegraphics[width=\linewidth]{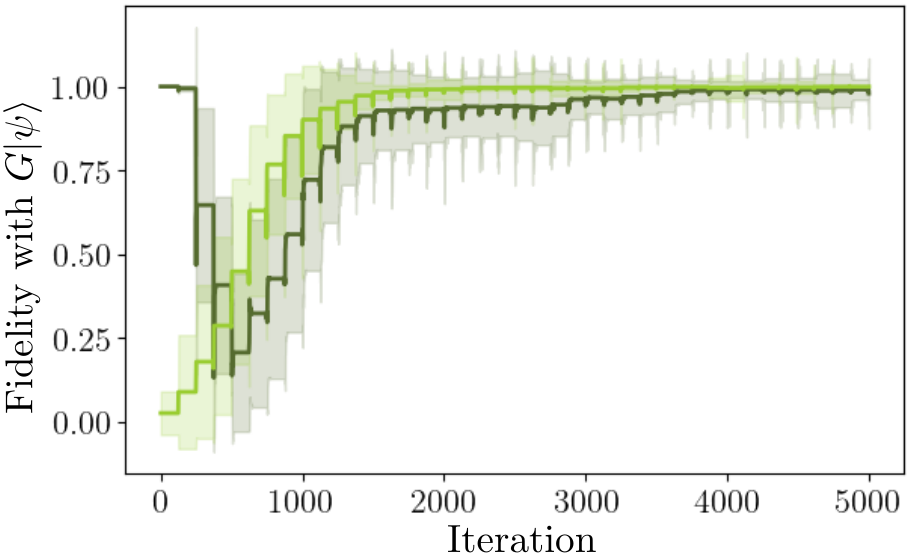}
    \caption{\textit{Average fidelity of Gauss law operator $G_l$ across each site during the optimization.} 
    For the ZZ ansatz we plot how the fidelity varies for optimization using SLSQP for the Hamiltonian with $J=m=\mu=1$ across 100 runs and we include the standard deviation at each iteration. 
    In dark green, we initialize in the gauge-invariant subspace by initialising the angles to $\pi$, whereas in light green this choice is random.
    Beginning in the subspace, within a few iterations, the VQE explores states outside of the space to then return, arriving on the state that minimizes the cost function.
    When the optimization begins outside of this subspace, the VQE nonetheless converges upon the state within the physical subspace.
    }
    \label{fig:fidelity-zz}
\end{figure}

\section{Results}

In this section we present the findings of numerical experiments simulating VQE and experiments using IBM's quantum hardware. 
First we motivate the choice of ansatz circuits. We then consider questions of scalability and classical simulability.
We go on to show results for ground state estimation and static string breaking, while considering fixed numbers of shots and the presence of noise. 
The numerical experiments are verified using tensor network calculations, and in doing this we find cases where the VQE can identify the ground state, while the TN converges to local minima, so that those TN simulations require the addition of penalty terms.

Classical simulation of the VQE was performed using Pennylane~\cite{bergholm2022pennylaneautomaticdifferentiationhybrid}, and we have used Qiskit~\cite{javadiabhari2024quantumcomputingqiskit} to interact with IBM's quantum platform.
We use the classical optimizers SLSQP \cite{nocedal2006numerical} and SPSA \cite{spall1998overview}, where we find SLSQP to be more performant. 
The estimated energies are compared against TN calculations that use the ITensor library,  where we run a standard two-site algorithm updating two neighboring tensors~\cite{Fishman_2022,Fishman2022b}.

\subsection{Choice of ansatz}

We consider two forms of ansatz circuit seen in Fig.~\ref{fig:ansatz} that we label gauge-invariant (GI) and one based on $Z$-interaction gates (ZZ) that we go on to explain. 
In the figure, the red gates represent state initialization and in green we include one of the parametrized repeating layers.
For the simulations we are limited to three plaquettes (that amounts to 18 qubits) and use no more than three layers, 
{to find good approximations of the ground state.} 
One layer of the GI ansatz for one plaquette requires $17$ parameters, while for ZZ it is $21$.
Qubits are assigned starting from the bottom left corner of the ladder (Fig.~\ref{fig:lgt-diagram}) and then traversing up then left. 
In this way, the qubit assignments run as matter- link- matter-qubits, then link- link-qubits. 

The state initialization step selects a Gauss law sector, which corresponds to different configurations of static charges. 
In Fig.~\ref{fig:ansatz}a the initial state is the zero charge sector, where $G_l\ket{\psi}=\ket{\psi}$, $\forall l$. In the ansatz, the alternating $X$ gates take into account the staggering. 
Common to all of the initializations to Gauss law sectors is the Hadamard gate $H$ acting on each of the link qubits.

Not including the state initialization step would lead to slow convergence of the VQE or trapping in local minima. 
For the GI ansatz, it is necessary to initialize in the correct Gauss law sector. 
In Fig.~\ref{fig:ansatz}a, we have initialized to the zero-charge sector.
The alternating $X$ gates take into account the staggered fermions.
Placing charges in the theory changes the Gauss law sector.
Practically, this amounts to adding an extra $X$ gate to the matter qubit with the charge. 
For the ZZ ansatz we can initialize to any charge sector, so we choose that which uses the fewest gates, as can be seen in the simpler state initialization step in Fig.~\ref{fig:ansatz}b.

Then for the evolution, enforcing gauge invariance can be difficult on a quantum computer since introducing large penalty terms can lead to barren plateaus.
Preserving gauge invariance is also difficult, since errors in the system evolution can push the state outside of the physical Hilbert space. 
One way to enforce and preserve gauge invariance is by constructing a gauge-preserving ansatz~\cite{mazzola2021gauge, alexandrou2025realizingstringbreakingdynamics}, and shown in Fig.~\ref{fig:ansatz}a. 
The GI ansatz is a Hamiltonian variational ansatz \cite{Park2024hamiltonian}, where the parametrized unitaries are each of the terms from the Hamiltonian Eq.~\eqref{eq:H}. 
Upon transforming Eq.~\eqref{eq:H_m} to spin degrees of freedom using the Jordan-Wigner transformation acting on the relevant matter and link qubits, the kinetic term introduces the unitaries proportional to, e.g., $\exp{(-i\theta/2 X\!\otimes\!Z\!\otimes\!Z\!\otimes\!Y)}$.

We also consider a simpler ansatz, that is presented in Fig.~\ref{fig:ansatz}b and we refer to as ZZ. 
In this ansatz, multi-qubit $Z$ rotation gates such as $R_{ZZZ}$ and $R_{ZZZZ}$ are placed where three- and four-body interaction terms appear in the Hamiltonian, i.e. the gauge part in Eq.~\eqref{eq:H_g} has 4-body plaquette interactions and the kinetic part of the Hamiltonian in Eq.~\eqref{eq:H_m} introduces 3-body interactions along the edges connecting matter sites (see Fig.~\ref{fig:lgt-diagram}).
Link qubits are acted on by $X$ rotations for the gauge field, and the matter qubits have the freedom of $X$, $Y$, $Z$ rotations. 
In measurement-based quantum computing (MBQC)~\cite{mbqc_def}, such multi-qubit $Z$ rotations can be implemented using one ancillary qubit connected to the relevant $n$ qubits~\cite{qin2024applicability}, so this ansatz probes the use of such interactions.
The choice of simpler gates also makes this ansatz more hardware-efficient. 

\begin{figure}
    \centering
    \includegraphics[trim={0cm 0.65cm 0cm 0.6cm},clip,width=\linewidth]{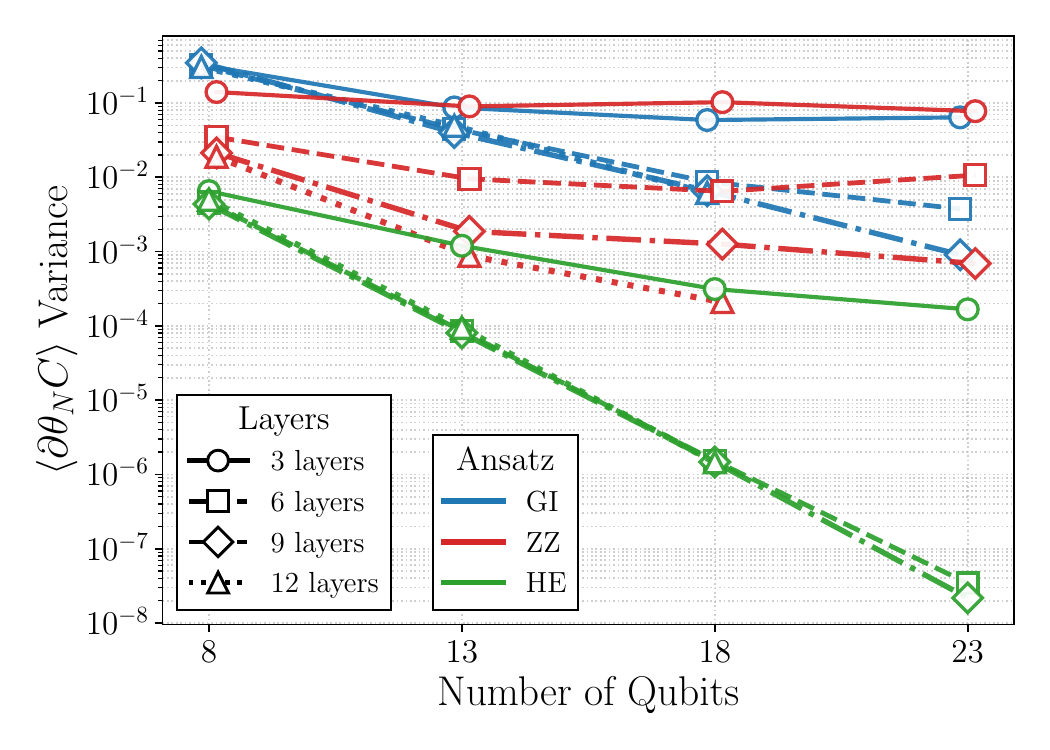}
    \caption{\textit{Scaling of the variance with system size across ansätze and numbers of layers.}
    Using the GI, ZZ, and strongly-entangling HE ansatz, across 3, 6, 9, and 12 layers, averaged over 500 samples (140 for the 23 qubit case).  
    The generic HE ansatz consists of layers of single qubit rotation gates followed by entangling CNOT gates between pairs of qubits. 
    As the system size is increased, we see a favorable scaling in the variance, projecting that gradients remain large enough for training systems of more qubits, towards the NISQ regime. }
    \label{fig:variances}
\end{figure}

Furthermore, this ansatz is gauge invariant when we initialize close to the angle $\theta\approx\pi$ for all the parametrized gates, allowing relaxation beyond the gauge-invariant subspace, as seen in Fig.~\ref{fig:fidelity-zz}.
Here we plot the fidelity of the resultant state for $100$ samples at each stage of the optimization with the Gauss law operator $G_l$ (\ref{eq:gauss_law}) to check for Gauss law violations.
The typical number of iterations to arrive at the physical subspace is seen through the standard deviation. 
This behavior can be seen to be generic, providing an idea of how the optimization explores the physical Hilbert space and its boundary. 
While the GI ansatz remains gauge invariant throughout the noiseless optimization, we expect the ZZ ansatz would require the enforcement of Gauss law through penalty terms. 
However, regardless of the Gauss law sector initialization, the VQE can learn the gauge-invariant ground state.

{In the investigations we find that using shallow depth circuits (up to three layers), the VQE with either ansatz is able to arrive at good approximations of the ground state, while for the ZZ ansatz, it is able to learn the correct gauge degrees of freedom in the absence of penalty terms.
To analyse the phenomena of barren plateaus we consider the scaling of the variance of the cost function with increasing the numbers of plaquettes (up to 3, and some instances of 4) and numbers of layers (up to 12). 
We also consider the scaling of the dynamical Lie algebra (DLA) with increasing number of qubits. 
}

In Fig.~\ref{fig:variances} we plot {the variance of the cost function, \eqref{eq:H}, with parameters $J=m=\mu=1$, averaged over $500$ samples ($140$ for the 4 plaquette case). 
These are compared to a hardware efficient (HE) ansatz consisting of layers of rotation gates parametrized by three angles and entangling CNOT gates (for one plaquette each layer has $24$ parameters). 
Both ansatz shows decay of gradients, while the ZZ ansatz may manifest larger gradients with increasing number of qubits.
The trend suggests that for the numbers of qubits that go beyond classical simulation and within the regime of NISQ, the manifested variances are large enough for training. 
That is, for VQE applied to this problem, we obtain accurate approximations of the ground state with shallow circuits, and the decay of variances is slow enough, allowing for deeper circuits and larger system sizes.   
With more layers the VQE becomes un-trainable, although arguably in the context of NISQ, for deep circuits, sources of errors would dominate before the issue of barren plateaus can arise. 
}

\begin{figure}
    \centering
    \includegraphics[width=\linewidth]{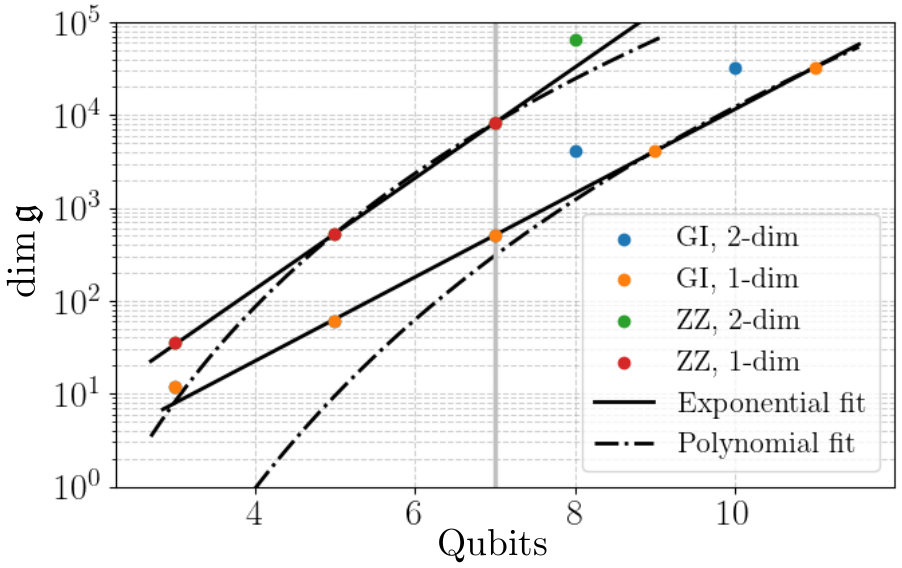}
    \caption{\textit{Scaling of the dimension of the DLA.} For increasing number of qubits of the GI and ZZ ansatz and the one- and two-dimensional LGT.
    Here we include an exponential and polynomial fit for the one-dimensional cases. 
    For the GI ansatz the exponential fitting is given by $\dim(\mathfrak g) \approx \num{3.49e-1} 2^{1.50n}$, where $n$ is the number of qubits, and the mean squared error (MSE) $10.56$, while the polynomial fit is $\dim(\mathfrak g) \approx \num{5.71e-7} n^{10.33}$ with MSE $8550.98$.     
    Whereas for the ZZ ansatz the exponential fit is $\dim(\mathfrak g) \approx \num{5.47e-1} 2^{1.98n}$  with MSE $1.63$, and then the polynomial fit is $\dim(\mathfrak g) \approx \num{1.04e-3} n^{8.17}$ with MSE $258.65$. 
    Up to $7$ qubits, the one- and two-dimensional cases are the same, as the 8\textsuperscript{th} qubit introduces a plaquette term. 
    With this additional term, the DLA is larger than for the one-dimensional case. 
    Also, the ZZ ansatz has a consistently larger DLA since this ansatz spans states outside of the physical subspace. 
    }
    \label{fig:dla_scaling}
\end{figure}

{To consider the scaling of the dimension of the DLA, $\dim\mathfrak{g}$, we take the un-physical scenario of adding one qubit to the theory at a time and thus its corresponding parameterised gates. 
This is because the compute required to scale as plaquettes is too high. 
These results are shown in Fig. \ref{fig:dla_scaling}, where we see that the one-dimensional problem of both GI and ZZ ansatz show exponential scaling. 
Up to $7$ qubits, the two-leg ladder is the same as the one-dimensional chain, so we cannot use these points to compute the scaling, but we see a larger $\dim\mathfrak{g}$ here.}


\begin{figure*}
    \centering
    \includegraphics[trim={0.5cm 0.3cm 0.5cm 3.5cm},clip,width=\linewidth]{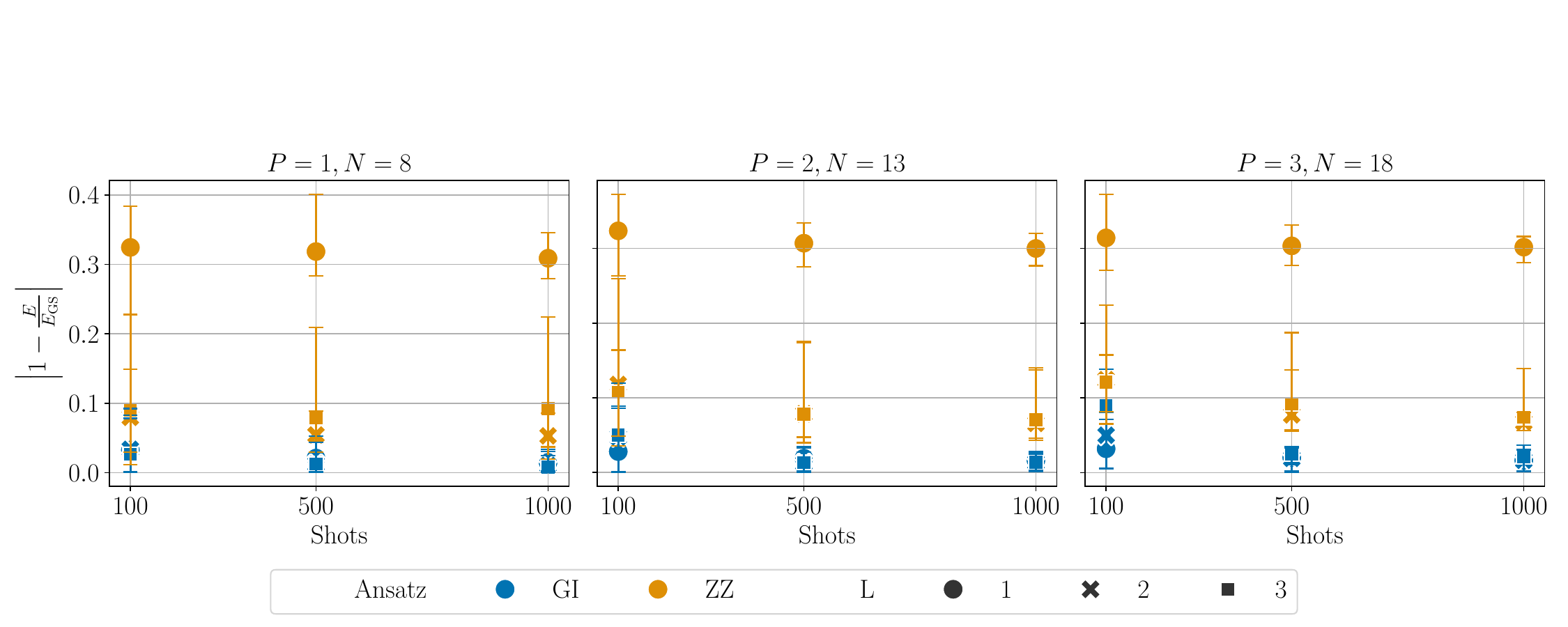}
    \caption{\textit{Performance of the VQE}. We variationally estimate the ground state of the Hamiltonian in Eq.~\eqref{eq:H} with $J=3$, $m=1$ and $\mu=2$, while increasing the number of plaquettes $P$, i.e., qubits $N$, numbers of ansatz layers $L$, and number of shots.
    On the $y$-axis we plot the energy relative error ratio with the ground state energy $E_{GS}$ obtained using tensor network methods. 
    As the system size is increased, the behavior of the VQE remains stable, which is a good indication for using the VQE for larger system sizes.
    For each instance we have taken the average of $20$ runs and used the SPSA optimizer. }
    \label{fig:varying_shots}
\end{figure*}

{Thus the $\mathbb{Z}_2$ LGT can be interesting for larger scale studies using VQEs and LGTs generally seem suited to VQE. 
For the average case it is not known how best to initialize in the cost landscape, while in LGT this is answered for by Gauss law sectors and the physical subspace. 
The VQE also leverages symmetries of the problem, and in this setting this is the gauge-invariant subspace. 
A problem is classically simulable if this subspace is polynomially sized, but we know the overall gauge-invariant subspace to be exponential in size~\cite{Buividovich_2009, Tagliacozzo2014}.}

\subsection{Ground state estimation}
\label{sec:results-gse}

We use VQE to find the ground state over a range of Hamiltonian parameters, and can be seen in Table \ref{tab:gs}. 
The limit $J\rightarrow0$ is known as the non-interacting limit, and with stronger interactions $J$, the ZZ ansatz is seen to perform less well. 
This performance is improved by including penalty terms, with $V \sim J$. 
We consider system sizes up to three plaquettes, and simulations with up to three repeating layers (see Fig.~\ref{fig:ansatz} for a schematic of an ansatz layer). 
The obtained ground state energy is compared to the tensor network (TN) calculation.

We show results for varying the numbers of shots, system size and numbers of layers for an instance of a Hamiltonian configuration in Fig.~\ref{fig:varying_shots}.
Here we use $J=3$, $m=1$, $\mu=2$, up to $3$ layers and for system sizes up to $3$ plaquettes. 
We find that with increasing system size the performance of the VQE remains roughly constant. 
For lower numbers of shots, the ansatz performance is not increased with increasing layers. 
We also see that one layer of the GI ansatz can obtain good convergence, so the ansatz is already quite expressive for this case. 

\begin{figure}
    \centering
    \includegraphics[width=\linewidth]{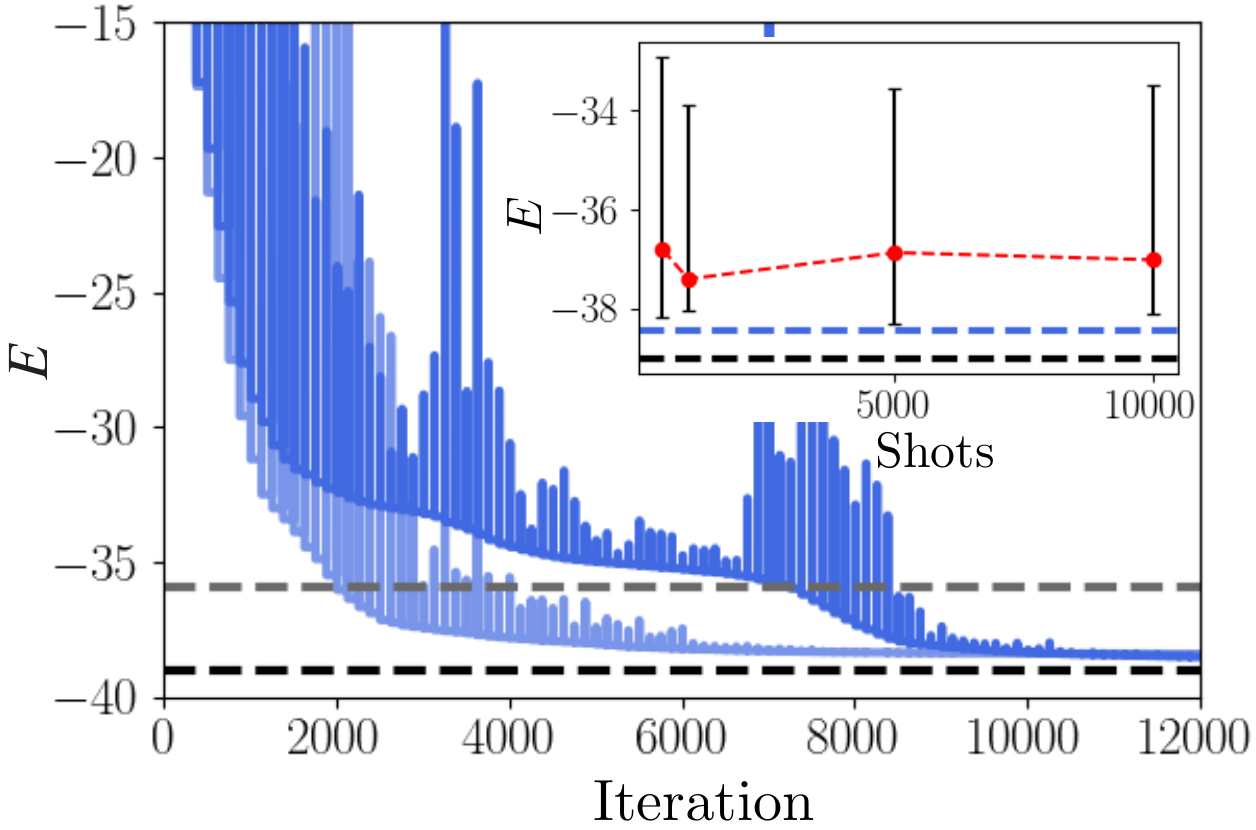}
    \caption{\textit{VQE optimization of the ground state.} Here we consider the Hamiltonian in Eq.~\eqref{eq:H} with $J=5$, $m=1$, $\mu=2$ and charges placed at $(0,0)$ and $(2,0)$.
    The main curves show two instances of the full state-vector simulation for a single run of VQE optimized using SLSQP. 
    The bold curve is a generic instance, while the translucent curve shows the case where the VQE escapes the local minima. 
    The grey dashed line is the ground state energy ($E=-35.954$) obtained using TN methods and we can see the simulation is stuck in the same local minima. 
    The black dashed line is the reference TN simulation ($E=-39.019$) with the addition of a penalty term $V=100$ that we compare our results with. 
    Without the addition of penalty terms, the VQE is successful in converging to the ground state. 
    In the inset we show results also for fixed numbers of shots $[500, 1000, 5000, 10000]$ averaged over $15$ runs. 
    In the software, rather optimizing with SPSA enables us to consider fixed numbers of shots.
    It seems the final energy does not depend on the number of shots. 
    The black dashed line is the ground state obtained using tensor networks, while the blue dashed line is the average of the SLSQP runs. 
    We note that on average, with this configuration SPSA attains values higher than if we optimized using SLSQP. 
    }
    \label{fig:groundstate_shots}
\end{figure}

In Fig.~\ref{fig:groundstate_shots}, we plot the optimization to find the ground state energy of the Hamiltonian in Eq.~\eqref{eq:H} with $J=5$, $m=1$, $\mu=2$ and charges placed at $(0,0)$ and $(0,2)$.  
In this case the corresponding TN calculation is stuck in a local minima, with this value plotted as a horizontal line. 

In the main plot of Fig.~\ref{fig:groundstate_shots} we plot two instances of state vector simulation of the VQE with the GI ansatz. 
In one of these instances the simulation is able to converge towards the ground state energy, while the other is an example where the VQE escapes a local minimum -- the local minima that the TN simulation converges to.
In the subfigure, we plot results for varying numbers of shots $[500, 1000, 5000, 10000]$, and we find that the behavior of the VQE is less affected by increasing the number of shots.
We note here that experiments with fixed numbers of shots use the SPSA optimizer, which we find to be less performant than SLSQP in this setting. 

\subsection{String breaking}

\begin{figure}
    \centering
    \includegraphics[trim={0cm 0cm 0cm 0cm},clip,width=\linewidth]{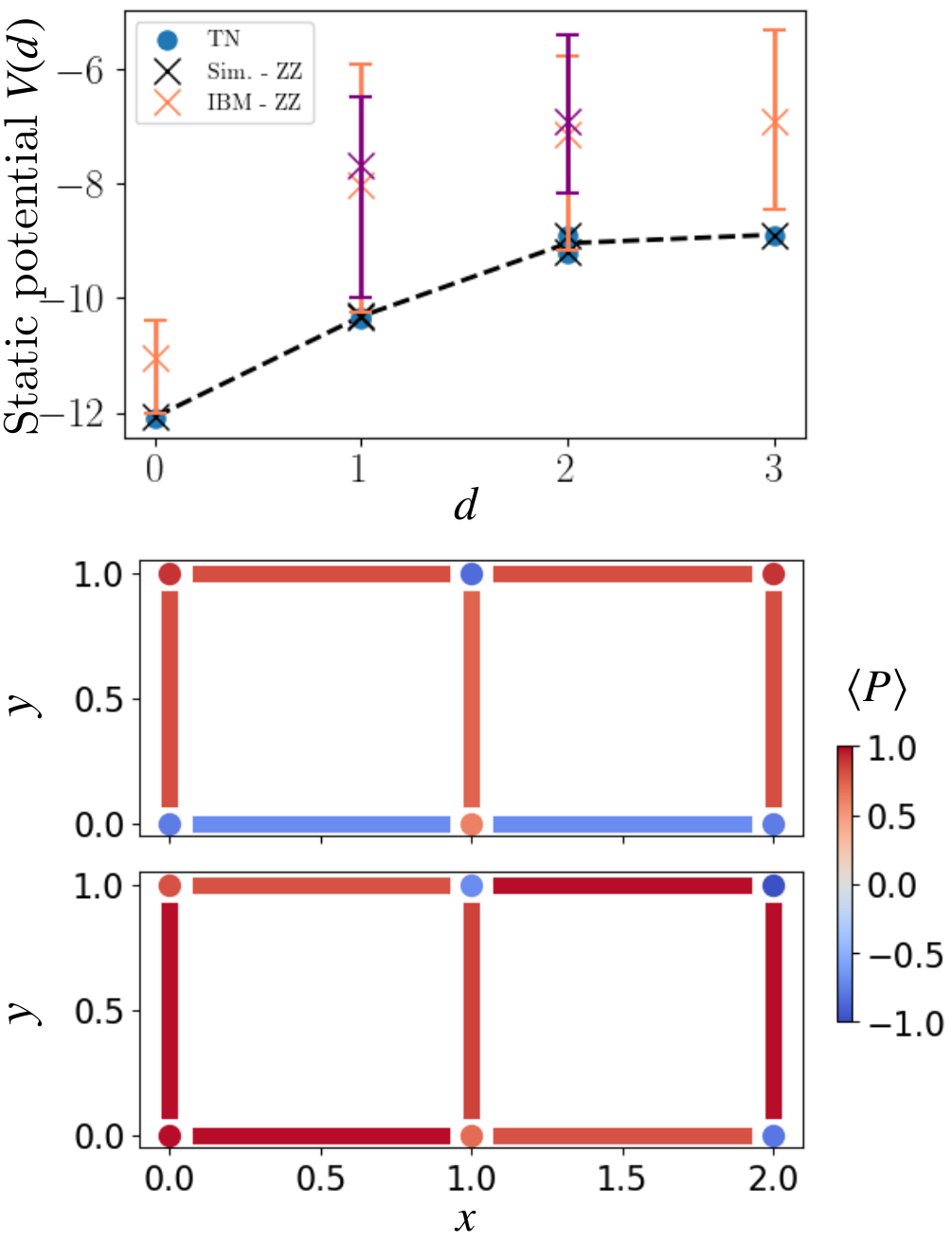}
    \caption{\textit{Observation of string breaking on quantum hardware.}
    Here we consider the $P=2$ plaquette $\mathbb{Z}_2$ LGT with Hamiltonian parameters $J=0.5$, $m=1.5$, and $\mu=1$, which has a string breaking transition at $d=2$. 
    We show results for the ground state energy with no charges ($d=0$) and pairs of charges with one fixed at $(0,0)$ and another at different lattice sites, with spacing $d$ from $(0,0)$.
    There are multiple values for $V(d)$ at $d=1,2$ due to the boundary conditions.
    The effect is more pronounced at $d=2$ where at $(1,1)$ there is one boundary condition and broken flux string, while at $(0,2)$ there are two boundary conditions and a flux string. 
    The black line is the mean of the $V(d)$, which shows the expected behavior of $V(d)$ -- this is a linearly increasing potential, up to the point where string breaking occurs and it becomes energetically favorable to break the string, thus $V(d)$ becomes constant.
    We plot the results for the TN simulation, one run of the state vector simulation of the VQE, and state preparation using the optimized rotation angles from the VQE on \texttt{ibm\_marrakesh}. 
    The IBM results summarize $20$ runs on the device, we plot the average value and error bars.
    The VQE is performed using three layers of the ZZ ansatz. 
    The TN simulations and the VQE simulations have strong agreement, which means that the resulting angle parameters for the state preparation using IBM are close to the ground state. 
    We see the qualitative behavior of string breaking and the two phases (existence of string and string-broken). 
    In the bottom figure we depict configurations of the lattice at criticality, between $d=2$ and $d=3$.
    The expectation values $\langle P\rangle$ correspond to the observable $\langle Z\rangle$ on the matter sites and $\langle X\rangle$ along the links. 
    With charges placed between $(0,0)$ and $(2,0)$ ($d=2$), we see the presence of a flux string, then extending this second charge to $(2,1)$ breaks the string, and we can see the values of $\langle Z\rangle$ becomes flipped, indicating the creation of a charge. }
    \label{fig:ibm-results-2}
\end{figure}

We also use the VQE to investigate static string breaking. 
Here we consider the two plaquette lattice, where we introduce pairs of charges within the theory and increase the distance between them. 
In the Hamiltonian of Eq.~\eqref{eq:H}, this surmounts to introducing site-dependent terms proportional to Eq.~\eqref{eq:gauss_law} that take the opposite sign in the coefficient.
The notion of charge is thus charged or uncharged. 
This can be observed for certain parameter regimes of Eq.~\eqref{eq:H}, where the string breaking critical distance is on the order of the lattice size.
This is true for the Hamiltonian Eq.~\eqref{eq:H} with $J=0.5$, $m=1.5$, and $\mu=1$.
We present these results in Fig.~\ref{fig:ibm-results-2}.

String breaking can be seen through the static potential $V(d)$, where $V(d)=E_0 - E_c$, and $E_c$ is the ground state with charges.   
The lattice distance $d$ is taken with reference to the fermion at $(0,0)$, so $d$ counts the number of links between this and the second fermion. 
This means that for distances $d=1,2$ there are two possible fermion sites to place the static charge, whereas for $d=3$ the only choice is site $(3,1)$. 
The values of $V(d)$ for $d=1,2$ do not agree across all the simulations due to boundary conditions, with this effect being more pronounced at $d=2$.
Furthermore, the charge at $(2,0)$ still has the presence of a flux string, while for $(1,1)$ we find the string vanishes. 
The effects of left and right boundary conditions can be mitigated using longer ladder geometries, however in the $y$-direction the boundary is sharp so in general $V(d)$ will not agree across the pairs.

Here we use three layers of the more hardware efficient ZZ ansatz, so we can go on to implement this with IBMQ. 
Running the entire VQE would be too costly, so we consider the state preparation step that is required at every iteration. 
For the simulations we perform full state vector simulation to obtain the best convergence, although, as seen in Fig~\ref{fig:varying_shots}, we do not expect using a fixed number of shots to have a tangible effect.
We perform one run of 100 iterations of the simulated VQE using Qiskit and then perform the final optimization loop iteration with those optimized angles on the quantum hardware. 
Results were obtained using IBM’s 156 qubit Heron r2 processor, and the QPU we use is \texttt{ibm\textunderscore marrakesh}.

The black dashed line plots the average value of $V(d)$, where we see a initial linear increase in $V(d)$ as the charges are separated and the flux string grows, then for some $2<d<3$, it becomes energetically favorable to break the string by forming two new charges. 
This behavior is seen in the lattice configuration plots of Fig.~\ref{fig:ibm-results-2}. 
Here we have measured the observable $\langle Z \rangle$ on the matter qubits and $\langle X \rangle$ on the link qubits. 
The observable $\langle Z \rangle$ tells us the sign of the staggering and where $\langle X \rangle = -1 $ indicates flux. 
In the upper diagram, we see the presence of a string when a charge is placed at $(2,0)$. 
In the lower diagram, where a charge is placed at $(2,1)$, we see the effects of string breaking, as the flux string is no longer present, and the opposite sign of $\langle Z \rangle$, which indicates charge creation.

Fig.~\ref{fig:ibm-results-2} also contains results obtained using the QPU \texttt{ibm\textunderscore marrakesh}. 
The experiments use $5000$ shots, and each data point is the average of $20$ runs on the device. 
The obtained ground state energies are presented in Table \ref{tab:IBM_data}.
We use the built-in Qiskit SDK error mitigation of zero-noise extrapolation (ZNE) and Twirled Readout Error eXtinction (TREX) \cite{vandenBerg2022Modelfree}, and for the transpiler we use ``high optimization'', so that the process to select the best qubits in the graph is run with greater effort and more trials. 
Idle qubits in a circuit can invite coherent errors through unwanted interactions between them, then it is beneficial to use dynamical decoupling, which works by inserting pulse sequences that are identity operations on the idling qubits to cancel out this effect. 
Operations in our circuits seem to be packed quite densely, so dynamical decoupling could be ineffective and thus have an adverse effect due to imperfections in the pulse sequences. 
We consider turning this off across a comparative number of runs, however we find the energy can be lower than the simulated value.

For the IBM experiments, the best state preparation performance is in the case without charges. 
The qualitative form of the string-breaking transition can be discerned in the data, however the range of obtained energies is quite large, yet evenly spread. 
It can be seen that for the charge placed at $(1,1)$ (purple error bar) and $(2,1)$ appear to have the same spread of static potential, which is also true of the simulated data as here the string has already broken. 
We disqualify shot noise as having a strong influence on the data, as repeating the experiments with $100000$ shots leads to similar behavior. 
It is not possible to distinguish the individual runs based on the ZNE data at different noise levels, which suggests some further optimization is occurring in the background.
The issue can lie in using ``high optimization", while choosing gates based on the error profile can improve average fidelity, it would introduce greater run-to-run variability. 
Even for consecutive runs on the device, the region of the device chosen can vary greatly, and perhaps the source of error being optimized for is not as relevant for these experiments. 
Fixing the layout of the qubits would make the average performance worse, but would make the data and ZNE curves more stable, and given the number of runs ($\sim 20$) this may have been the better choice.

Overall the numbers of interacting gates in the two plaquette case seems manageable for using IBM quantum hardware. 
The GI ansatz has comparable performance with fewer numbers of layers, (see Fig.~\ref{fig:varying_shots}) so it would be interesting to perform similar analysis with one layer of the GI ansatz. 
The degree to which the ZZ ansatz is hardware efficient can also be determined in this way. 
It would also be important to consider one connected subspace of the hardware and perform more runs.

Initially we considered three plaquettes of the LGT, where a string breaking transition can be seen for $J=5$, $m=1$ and $\mu=3$, results for these are shown in Appendix \ref{app:string_breaking} and results for the case $\mu=2$ are seen in the previous section. 
Between the values $\mu=2$ and $\mu=3$, there is a transition such that string breaking occurs at the lattice spacing $d=3$. 
However, when performing the optimization with Qiskit, convergence could not be achieved in a reasonable time so that the state being prepared on the device was not the optimal. 
The results indicate that we should consider smaller system sizes, so we looked at two plaquettes. 
Here we find that the resulting state we prepare on the quantum device is already close to the ground state. 

\section{Discussion}

In this work we use a VQE to investigate a $\mathbb{Z}_2$ LGT coupled to Kogut-Susskind fermions on a two-leg ladder lattice. 
We are able to recover ground state properties and aspects of static string breaking using this technique. 
In doing so, we find that $\mathbb{Z}_2$ LGTs have features that suit investigation using VQEs, 
{such as good approximations to the ground state using shallow circuits and the ansatz circuits themselves avoid barren plateaus.}

One way to avoid barren plateaus is to use smart initialization strategies~\cite{Puig2025}; however a general strategy to identify these or to avoid local minima is not known.
For LGTs on the other hand, we always initialize in some Gauss law sector and would know beforehand which to initialize in. 
Further, checking for Gauss law violations can alert as to whether the solution is a local minimum. 

Another way is to constrain the searchable Hilbert space. 
This is done through the choice of ansatz, which is commonly designed to respect the symmetries of the problem. 
Solutions of LGTs are constrained to the gauge-invariant subspace of Hilbert space, and here we present an ansatz of gauge-invariant Hamiltonian terms that remains in this space. 
When the subspace spanned by the ansatz is polynomial in size, the problem can be classically simulated. 
All the while, the full space of gauge-invariant states is exponentially large~\cite{Buividovich_2009, Tagliacozzo2014}.

In this work we present two ansatz circuits, one that is constructed of gauge-invariant Hamiltonian terms, this we refer to as GI, and another simpler ansatz called ZZ, that is composed of multi-qubit $Z$ rotation gates. 
Despite not being fully gauge-invariant, nor taking into account qubit interactions from the Jordan-Wigner transformation, the ZZ ansatz is effective for the convergence of the VQE to the ground state.
Further, by tracking the fidelity with the Gauss law operator, we see how this ansatz explores Hilbert space, to arrive in the gauge-invariant subspace.  
As such multi-qubit $Z$ interactions can be a simple, MBQC hardware-efficient, way for the ansatz to capture multi-qubit interactions.  
Both of these ansatz demonstrate good scaling in the variance of the cost function with increasing system size for reasonable numbers of qubits and circuit depths.
We also come across instances where the tensor network simulation is stuck in local minima, while the VQE is able to learn the correct gauge degrees of freedom. 

Fundamentally, VQE cannot be implemented in ideal conditions, so considering the effects of scaling with system size, number of shots and incorporating the effects of noise is vital for making such a determination.  
We find that even with low numbers of shots in the simulations, $\sim 500$, convergence to the ground state is accurate and the resulting state is gauge-invariant. 
We also find that when scaling with system size, this behavior remains constant, which indicates that for larger systems placed on quantum devices, the VQE could behave as we expect. 

Preliminary studies of the effects of noise are performed using IBM's quantum platform. 
Here we obtain results using the ZZ ansatz for a string breaking transition of a two plaquette $\mathbb{Z}_2$ LGT.
Since these circuits decompose to become quite deep, we are surprised to find that with the default error mitigation, and not fine-tuning the device, we can arrive at low energy states and distinguish the string breaking transition. 
In a limited capacity, we explore this further, by performing experiments with a three plaquette $\mathbb{Z}_2$ LGT.
Here we can also see some hallmarks of the transition. 
We would look to see if considering a fixed layout of a connected region of the hardware would improve results, and the best error metric to choose qubits.

In future work we would look to perform more experiments using quantum hardware. 
For smaller system sizes, it would be useful to perform some rounds of optimization in the VQE to see whether the resulting state remains gauge-invariant, and extending to observations of the string-breaking transition, where it would be interesting to see if this persists. 
Furthermore, erroneous results can be distinguished if gauge-violations are present, which is a diagnostic we should next time include. 
We are also interested in the effects of small penalty terms on the cost landscape, since small terms in a region $\sim J$ can benefit the convergence of the $ZZ$ ansatz. 
The small penalty terms may also mitigate noise that can push the state out of the gauge-invariant subspace. 
Larger ladder lattice sizes would mitigate the boundary effects and we have performed preliminary studies involving square lattices, showing that we can extend this approach.

The behavior of the GI ansatz using current hardware versus the more hardware-efficient ZZ ansatz can also be analyzed, which would indicate the practicality of scaling such a VQE in the NISQ era. 
For those studies it would be beneficial to work more closely with the quantum hardware provider.
It would also be beneficial to extend these studies to other platforms too.
For instance the long-range multi-qubit interaction gates of the two ansatz may better suit ion-trap~\cite{Molmer2000} or neutral atom based~\cite{baker2021exploitinglongdistanceinteractionstolerating} quantum hardware. 
It is also interesting to explore applying the ZZ ansatz within the framework of measurement-based VQE~\cite{Ferguson2021VQE}, and explore the effectivity of having multi-qubit $Z$ rotation gates in the ansatz where the problem Hamiltonian has many-qubit terms. 

Another avenue to explore is to ask why it might be the case that this $\mathbb{Z}_2$ LGT is suited to VQE. 
Where fault-tolerant quantum algorithms might succeed and variational approaches might fail is within the circuit unitaries. 
Fault-tolerant quantum algorithms rely on precisely engineered unitary gates, while VQEs have much redundancy here. 
Then perhaps the $\mathbb{Z}_2$ LGT benefits from being a theory made of simple Pauli operators, that is naturally encoded in terms of qubits. 
It would be important to extend these studies to other gauge groups, and to further explore the relevance of gauge symmetries.

\begin{acknowledgments}
    This work has been supported with funds from the Ministry of Science, Research and Culture of the State of Brandenburg within the Center for Quantum Technology and Applications (CQTA). It also received support from the 
    Federal Ministry for Economic Affairs and Climate Action (project Qompiler, grant No: 01MQ22005A), the DFG (Emmy Noether programme, grant No. 418294583) and the Berlin Quantum Alliance.
\end{acknowledgments}

\FloatBarrier
\bibliography{references}

\appendix

\section{Ground state energies across parameter space}

In table \ref{tab:gs} we present some values for the ground state energy obtained using VQE across parameter space. 

\begin{table}[h]
\label{tab:gs}
\begin{tabular}{cccccccc}
\toprule
$J$ & $m$ & $\mu$ & GI & ZZ & TN \\
\midrule
0.01 & 0.01 & 0.01 & -2.05 & -2.05 & -2.05\\
 &  & 1.0 & -7.28 & -7.28 & -7.28\\
 &  & 10.0 & -70.05 & -70.06 & -70.06\\
  \cline{2-6}
 & 1.0 & 0.01 & -5.00 & -5.00 & -5.00\\
 & & 1.0 & -10.25 & -10.25 & -10.25\\
 &  & 10.0 & -73.03 & -73.03 & -73.03\\
  \cline{2-6}
 & 10.0 & 0.01 & -32.00 & -32.00 & -32.00\\
 &  & 1.0 & -37.25 & -37.25 & -37.25\\
 &  & 10.0 & -100.03 & -100.02 & -100.03\\
 \hline
1.0 & 0.01 & 0.01 & -6.46 & -6.06 & -6.46\\
 &  & 1.0 & -9.66 & -9.20 & -9.67\\
 &  & 10.0 & -70.40 & -70.24 & -70.40\\
 \cline{2-6}
 & 1.0 & 0.01 & -7.41 & -7.15 & -7.41\\
 &  & 1.0 & -11.77 & -11.64 & -11.77\\
 &  & 10.0 & -73.34 & -73.34 & -73.34\\
 \cline{2-6}
 & 10.0 & 0.01 & -32.35 & -32.28 & -32.35\\
 &  & 1.0 & -37.56 & -37.47 & -37.56\\
 &  & 10.0 & -100.20 & -100.20 & -100.2\\
  \hline
10.0 & 0.01 & 0.01 & -46.64 & -41.43 & -46.64\\
 &  & 1.0 & -47.71 & -45.22 & -47.77\\
 &  & 10.0 & -92.48 & -88.03 & -95.51\\
 \cline{2-6}
 & 1.0 & 0.01 & -46.75 & -44.30 & -46.75\\
 &  & 1.0 & -48.01 & -44.55 & -48.05\\
 &  & 10.0 & -94.39 & -92.77 & -94.43\\
 \cline{2-6}
 & 10.0 & 0.01 & -56.14 & -53.68 & -56.14\\
 &  & 1.0 & -59.25 & -56.92 & -59.30\\
 &  & 10.0 & -114.71 & -113.55 & -114.72\\
\bottomrule
\end{tabular}
\caption{Some values for the obtained ground state energy of Hamiltonian \eqref{eq:H} for the two plaquette LGT using GI and ZZ ansatz over a range of Hamiltonian parameters $J, m, \mu$.
The VQE used 3 layers of each ansatz circuit, and the SLSQP optimizer, with 200 as the maximum number of iterations.}
\end{table}

\begin{table}[b]
    \centering
    \begin{tabular}{cccccc}
    \toprule
       (0,0) & (0,1) & (1,0) & (1,1) & (2,0) & (2,1) \\
       \midrule
       -12.026 & -10.241 & -10.000 & -9.148 & -8.168 & -8.439 \\
-11.935 & -9.911 & -9.350 & -8.230 & -7.838 & -8.322 \\
-11.305 & -9.661 & -9.064 & -8.027 & -7.824 & -8.259 \\
-11.250 & -9.149 & -8.857 & -7.987 & -7.808 & -7.812 \\
-11.248 & -9.047 & -8.466 & -7.924 & -7.761 & -7.504 \\
-11.231 & -8.978 & -8.021 & -7.910 & -7.660 & -7.491 \\
-11.191 & -8.890 & -7.908 & -7.714 & -7.567 & -7.322 \\
-11.170 & -8.572 & -7.696 & -7.441 & -7.563 & -7.047 \\
-11.166 & -8.391 & -7.536 & -7.329 & -7.481 & -6.942 \\
-11.114 & -8.325 & -7.504 & -7.111 & -7.256 & -6.869 \\
-11.002 & -8.293 & -7.421 & -7.083 & -6.831 & -6.821 \\
-10.938 & -7.717 & -7.409 & -6.706 & -6.740 & -6.745 \\
-10.879 & -7.528 & -7.334 & -6.673 & -6.454 & -6.655 \\
-10.796 & -7.325 & -6.977 & -6.532 & -6.287 & -6.434 \\
-10.738 & -7.021 & -6.930 & -6.297 & -6.198 & -6.253 \\
-10.724 & -6.558 & -6.853 & -6.182 & -6.065 & -6.130 \\
-10.670 & -6.556 & -6.738 & -6.164 & -6.046 & -6.080 \\
-10.610 & -6.541 & -6.603 & -6.143 & -5.728 & -5.905 \\
-10.468 & -6.044 & -6.495 & -6.113 & -5.403 & -5.753 \\
          \midrule
          -12.093 & -10.346 & -10.340 & -9.217 & -8.898 & -8.900
    \end{tabular}
    \caption{Results from IBMQ in order of magnitude. The bottom row is the result of the simulated VQE. }
    \label{tab:IBM_data}
\end{table}

\section{Further details of IBMQ experiments}
\label{app:string_breaking}

\begin{figure}[]
    \centering
    \includegraphics[width=0.95\linewidth]{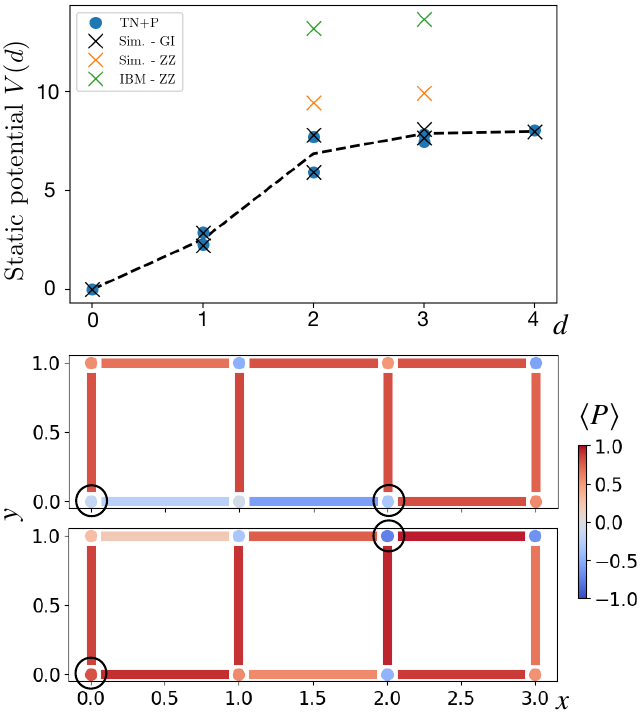}
    \caption{\textit{Observation of string breaking in the $\mathbb{Z}_2$ LGT on the $3$ plaquette ladder lattice.} The topmost figure plots the static potential $V(d)$ against increasing lattice spacing $d$ between the charges.
    The main plot shows results for one run of VQE using the GI ansatz.
    The lattice spacing is measured from $(0,0)$, and one of the charges is kept placed here. 
    Thus charges at $(0,1)$ and $(1,0)$ are $d=1$ apart, and for these equidistant pairs discernibility in their values of $V(d)$ is due to boundary effects.
    We also plot simulated values from one run using the ZZ ansatz. 
    Here the optimization could have run for more iterations, however due to the compute time it was terminated here. 
    The values of the ZZ ansatz are to be used in reference with the IBM results -- this is described in the main text. 
    Here we have also plotted the static potential of the state prepared on \texttt{ibm\textunderscore marrakesh} using the angles from this ZZ VQE simulation. 
    The best result of five consecutive runs was chosen. 
    In the bottom figure we depict configurations of the lattice at criticality, between $d=2$ and $d=3$.
    }
    \label{fig:string_breaking}
\end{figure}

Here we present those results for string breaking in the three plaquette LGT, which was discussed in the main text. 
We present the best run of five runs of the VQE for the Hamiltonian Eq.~\eqref{eq:H} with $J=5$, $m=1$, and $\mu=3$.
The case $\mu=2$ is discussed in the main text in Section \ref{sec:results-gse}. 
Between these values $\mu=2$ and $\mu=3$, there is a transition such that string breaking occurs at the lattice spacing $d=3$. 

The state we arrive at is not fully converged yet as this would require prolonged compute time, and for this we found that Qiskit was slow.
Yet, the state is gauge-invariant, so we use this. 
The resultant 141 angles are then used to prepare the state on \texttt{ibm\textunderscore marrakesh}. 

For the complicated circuits it is encouraging that the qualitative behavior of the VQE can be seen, but the results indicate that we should consider smaller system sizes. 
The initial angles for state preparation can be improved by using small penalty terms with the ZZ ansatz, since this would lead to faster convergence, so reduced compute time. 
This would also introduce another aspect of the experiments that requires fine-tuning, which was not possible with the limited access to hardware. 

\end{document}